\documentclass[12pt,a4paper]{article}

\usepackage[utf8]{inputenc}
\usepackage[T1]{fontenc}
\usepackage{mathptmx}
\usepackage[margin=1in]{geometry}
\usepackage{hyperref}
\usepackage{natbib}
\usepackage{url}

\hypersetup{
    colorlinks=true,
    linkcolor=blue,
    citecolor=blue,
    urlcolor=blue
}

\title{Large language models are not the problem}
\author{Hiranya V.\ Peiris\thanks{e-mail: \href{mailto:hiranya.peiris@ast.cam.ac.uk}{hiranya.peiris@ast.cam.ac.uk}}\\[6pt]
\textit{Institute of Astronomy, University of Cambridge, Cambridge, UK}}
\date{}

\begin{document}

\maketitle

\begin{abstract}
\noindent If a Large Language Model (LLM) can replicate your scientific contribution, the problem is not the LLM. What does it say about our field that so much of the anxiety about AI comes down to the fear that a machine could do what we do? Perhaps it says we should be doing something better.
\end{abstract}

\noindent\textit{Published in Nature Astronomy (2026).}\\
\noindent\textit{DOI: \href{https://doi.org/10.1038/s41550-026-02837-2}{10.1038/s41550-026-02837-2}}

\bigskip

\begin{quote}
``Replicants are like any other machine --- they're either a benefit or a hazard. If they're a benefit, it's not my problem.'' Rick Deckard, \textit{Blade Runner} (1982)
\end{quote}

\bigskip

In Ridley Scott's \textit{Blade Runner}, Deckard's job is tracking down rogue artificial beings --- `replicants' --- which have become nearly indistinguishable from humans. He delivers this line with the weary confidence of someone who thinks the question is settled. If the tool is a benefit, it's not his problem. Naturally, it becomes his problem almost immediately --- and the rest of the film is about what happens when the boundary between tool and agent turns out to be less clear than anyone had assumed.

There is currently a lively and occasionally breathless debate in the astrophysics community about large language models (LLMs). I have been following it with interest, not least because I work in the space that much of the anxiety is about: my research team builds generative models for galaxy populations and develops machine learning (ML) methods for large photometric surveys. We use ML not as a black box but as a creative instrument, and it lets us do things that were simply not possible until recently. I also regularly use LLMs directly myself, in ways I will describe below. I have not found that this makes me a worse scientist --- if anything, the opposite.

But this piece is not primarily about LLMs; it is about what the anxiety surrounding them reveals. The arrival of these tools has thrown into sharp relief a set of problems with scientific practice, standards, and incentive structures that existed long before anyone had heard of ChatGPT. The paper mills, the thin incremental work, the volume-over-quality culture, the erosion of deep domain knowledge --- none of these are products of AI. They are products of how we have chosen to organize and reward scientific work. LLMs merely make the consequences harder to ignore.

\section*{Privilege and practice}

The work we do as individual scientists is threaded into something larger: the story that humanity has been weaving about the Universe and our place within it. I feel that it is an extraordinary privilege to be paid to do work that brings me such joy. In turn I owe a great debt to society, which amidst many pressing priorities, chooses to support that work. My professional identity is shaped by personal formative experiences of scarcity, as well as gratitude towards mentors who gave me opportunities to contribute to this human enterprise. Research in astrophysics is funded by taxpayers, philanthropic foundations, and individual donors who choose to support this work because they believe it matters. That privilege is not permanent, and it is not unconditional. People do astrophysics for all sorts of reasons~\citep{Hogg2026}. Whatever the motivation, being paid to pursue it is a privilege. The obligation that comes with that privilege is to do it well. This perspective strongly informs the lens through which I view the recent conversations about LLM use.

I am from Generation X --- the generation that made the analogue-to-digital transition --- and so engaging with new technology has never been optional for me, and I have never wanted it to be. I have changed research focus repeatedly throughout my career, often learning alongside a new graduate student so that we approach the problem with the `beginner's mindset'. Most recently, when my team and I began working with machine learning, I again started as a complete novice. Each of these transitions of research focus required learning to think in a new way. None of them diminished the science --- they all made it better.

Alongside doing the best work I can, I have devoted considerable effort to working out how to become a better scientist, something I expect to be a lifelong practice. I believe everything can be improved. I believe I can learn from anyone. I reflect, regularly and deliberately, on how to pick better problems to work on, how to sharpen my thinking, how to communicate more clearly, how to mentor more effectively, and how to improve the way our community conducts itself. This is the lens through which I assess any new tool: does it help me do these things? If it does, I will use it. If it stops helping, I will set it aside.

\section*{How I actually use LLMs}

Much of the current debate conflates very different ways in which LLMs enter the research process. I find it helpful to be concrete about my own practice.

I have always sharpened my thinking by talking with people whose judgment I trust and who bring complementary perspectives. A good interlocutor asks the question you had not thought to ask or finds the crack in an argument you thought was airtight. Over a career built on crossing between disciplines I have learned that these conversations are most valuable when the other person sees the problem from a different perspective than my own.

I find that modern LLMs --- I use Anthropic's Claude --- can play a version of this role. Not as a substitute for human colleagues, but as a readily available first pass: a way to stress-test an argument, identify a gap in my reasoning, or rapidly explore unfamiliar technical territory before taking the refined version to a collaborator. But the principle is the same one that has guided me since my first collaborations: if a conversation helps me think more clearly, I will have it. Often just the act of having to explain my reasoning clearly to someone else is what clarifies it. None of this enters my analysis. It enters my thinking, which is where it belongs. Nobody asks you to publish the transcript of every conversation you had in the hallway.

Members of my research team also use AI-assisted coding tools. The resulting code is not trusted without validation and review --- exactly as we would treat code written by any team member. It is tested against known cases, reviewed by others, and committed to version-controlled repositories. The final, validated code is just as reproducible and inspectable as code typed entirely by hand. The concern that LLMs are ``designed to never give the same answer to the same prompt''~\citep{Hogg2026} is legitimate, but it only applies if the LLM output is the final product. When it is an intermediate step --- reviewed, validated, and frozen into a reproducible codebase --- the non-determinism of the LLM is no more scientifically relevant than the non-determinism of human thought.

Where I share the community's concern is with the third possibility: AI systems that design, execute, write up, and submit scientific projects with minimal human oversight. When no one fully understands the chain of reasoning from question to conclusion, the essential character of the scientific enterprise --- in which people take responsibility for claims about the natural world --- is at risk. That concern is genuine and serious. But the difficulty with the current debate is that much of the anxiety appropriate to this scenario is being projected onto the other two, where it does not belong.

\section*{What the anxiety reveals}

The worry that LLMs will flood the literature with low-quality work is not unfounded. But it is incomplete, in a way that matters.

The astrophysics literature had a quality problem long before LLMs arrived. The incentive structures of our profession --- publish or perish, citation metrics as proxy for impact, volume as proxy for productivity --- have been producing incremental, poorly checked, and sometimes wrong papers for decades. Peer review was already buckling under the load. Code was already going unvalidated. The paper mill is not an AI invention; it is a human one.

Underlying much of the current anxiety is an assumption, sometimes stated explicitly~\citep{Ting2025}, that ideas are cheap in astronomy and that the field is rate-limited by the time it takes to turn those ideas into papers. I think this gets the problem almost exactly backwards. The hardest part of being a scientist is not execution. It is working out which problems are important --- and, just as critically, learning which papers not to write. In my chapter for a recent book on doctoral research~\citep{Peiris2021}, I described this as one of the most important skills a scientist can develop: the discipline to resist the pressure to be constantly occupied with busywork, and to instead allow yourself the time --- including the time to be bored, to free-associate, to let your mind wander --- to identify questions genuinely worth spending years on. Some incremental work is a natural part of learning --- the problem is a system that continues to reward volume long after that stage. The flood of incremental papers in our literature is not evidence that we have too many good ideas and too little capacity to execute them. It is evidence that we have allowed execution to substitute for thought.

LLMs may accelerate this problem. They may make it easier to produce superficially competent work that does not make any lasting contribution. But it is a risk of degree, not of kind. The disease was already present. The question is whether the response should be to restrict the tool, or to finally address the underlying pathology. If the incentive structure of the field rewarded quality rather than quantity, many of the concerns about LLM-assisted paper production would lose their force.

There is a related anxiety that I think deserves particularly close examination: the worry that LLMs will make ``data science astrophysicists'' redundant~\citep{Hogg2026}. If an LLM can design a data analysis, write the code, run it, and draft the paper, what is left for the human? Taken at face value, I think this worry is actually a diagnostic --- not of the tool's power, but of a mode of work that has become common in parts of our field. If a practitioner's contribution can genuinely be replicated by a statistical process with no understanding of the underlying physics, then the activity was not sufficiently scientific to begin with.

The best work at the intersection of astrophysics and data science has always been deeply rooted in domain knowledge. When my team builds generative models for galaxy populations, the choices that matter are astrophysical in nature. They require the kind of integrated physical intuition that comes from years of working with both data and theory. What LLMs can replicate is the more mechanical mode of work: running an existing pipeline on a new dataset; tweaking hyperparameters; producing a paper whose contribution is essentially `we did the same thing as X but on data Y,' or `we added extra parameters to a model and fit to the same data.' That mode was already scientifically thin. This is not to say that all routine work is without value --- domain intuition often develops through the act of doing, and there are skills that cannot be acquired any other way. The question is whether the scientist is learning from the process or merely executing it. If LLMs expose the difference, it is not a crisis for astrophysics. It is a long-overdue reminder that domain knowledge is not an optional extra bolted onto statistical technique. It is the core of the scientific enterprise.

\section*{The promise we make to students}

When I think about what matters most in my own career, it is not the papers or the recognition. It is the people --- the mentors who saw something in me worth supporting when I had very little to show for myself, and the students and postdocs I have worked with since, watching them develop into scientists with their own voices and their own questions.

Taking on a PhD student is one of the most serious commitments in academic life. I do not say this lightly. You are undertaking to help another person develop into the best scientist they can be --- and I mean that in a broad sense. It is not just about technical training. It is about helping someone learn to think clearly, to ask good questions, to handle failure, to communicate with precision, to develop their own taste and judgment. It is about helping them grow as a person. My own early career experience taught me that the best training gives people real responsibility to tackle hard problems, alongside the support to navigate them.

But if we are serious about the primacy of people, we need to ask some uncomfortable questions about the shape of our community. The academic pyramid has grown far more broad-based~\citep{Metcalfe2008} in the nearly three decades since I was a student. The number of PhD studentships has increased dramatically; the number of faculty positions has not, and is unlikely to do so. The result is a system that takes in far more people than it can absorb, trains many of them inadequately, and then ejects them --- frequently after years of postdoctoral uncertainty --- into alternative careers that often come with job satisfaction and good remuneration, but perhaps could have been achieved with less time spent in positions of precarity.

This is not caused by LLMs; it is caused by the incentive structures we have built. A principal investigator with a large grant can recruit many students, and in the current system there is every reason to do so: more students means more papers, more citations, more evidence of `impact' at the next funding review. But the quality of mentoring necessarily suffers when it is spread too thin. An advisor with a large number of students cannot give each of them the deep, sustained, individual attention that genuine intellectual development requires. And when an advisor does not have to reflect deeply on the quality of their ideas --- because any passing notion can be farmed out to a student --- neither the student nor the science is well served. The student learns to execute rather than to think. The resulting papers are often incremental, because the ideas behind them were never properly scrutinized. The literature grows, but understanding does not.

When people worry that AI tools will turn PhD students into ``prompt engineers''~\citep{Trotta2025}, they are describing a degradation of training that was already well underway before the widespread use of LLMs. The tools have changed; the underlying dynamic --- students as instruments of productivity rather than people being developed --- has not. If we want to protect what matters about graduate education, the place to start is not with AI policy; it is with honest reflection on whether the current model of student supervision is serving the people it claims to serve.

\section*{The middle way}

The distinction I think is missing from the current debate is between automating astrophysics and augmenting astrophysicists. Automating astrophysics means building systems that do the science without us --- and that is genuinely concerning, not so much because the outputs might be wrong, but because it removes the human act of understanding that makes the whole enterprise worthwhile. Augmenting astrophysicists means giving people better tools to do their own thinking, coding, and analysis. This is what telescopes do. This is what computers do. This is what LLMs do --- when used by someone who understands the science and takes responsibility for the result.

The trust problem that LLMs raise --- how do we rely on opaque tools whose workings we do not fully understand? --- is real, but it is not new. When my team develops interpretable generative models, the `interpretable' is not decorative --- it is the scientific content. The community needs to be able to understand, validate, and criticize the tools it uses. This was true before LLMs arrived and will remain true long after the current debate has moved on.

What is needed is not a grand policy framework imposed from above, but a set of community norms adopted from within --- extensions of those we already have about reproducibility, transparency, and intellectual honesty --- that accommodate the reality of these tools: clear disclosure of LLM use, as we already disclose software dependencies; the expectation that authors can explain and defend any analysis in their papers, regardless of what tools assisted in producing it; and a shift away from volume metrics towards quality in hiring and promotion. None of this is revolutionary. These are the standards we should have been enforcing all along.

Deckard was wrong, in the end, not because the replicants were dangerous, but because he had misunderstood the nature of the problem. The question was never really about the machines. It was about what kind of society had produced them and why. The same is true here. LLMs are not a threat to astrophysics. But the mirror they hold up to our profession is not entirely flattering, and we would do well to look at it honestly. The problems it reveals --- the perverse incentives, the erosion of mentoring, the substitution of volume for thought --- are ours, not the machine's. Fixing them is our problem. It always was.

\section*{Competing interests}

The author declares no competing interests.

\end{document}